# Classification of Various Security Techniques in Databases and their Comparative Analysis

**Harshavardhan Kayarkar**
M.G.M's College of Engineering and Technology, Navi Mumbai, India
Email: hjkayarkar@gmail.com

**ABSTRACT:** Data security is one of the most crucial and a major challenge in the digital world. Security, privacy and integrity of data are demanded in every operation performed on internet. Whenever security of data is discussed, it is mostly in the context of secure transfer of data over the unreliable communication networks. But the security of the data in databases is also as important. In this paper we will be presenting some of the common security techniques for the data that can be implemented in fortifying and strengthening the databases.

## 1. INTRODUCTION

In this ever growing cyber world where millions and trillions of bytes of data is transferred everyday over the internet, the security of this data is a top priority and a major challenge.

Data security has become a necessity for every individual who is connected to internet and uses the internet for any purpose. It is a requirement that is a must in every aspect of the operation performed on the internet. Operations like online money transactions, transfer of sensitive information, web services, and numerous other operations need security of data. Along with these operations on the internet, data security is also essential and important in databases.

Databases are the storage areas where large amount information is stored. The nature of information stored varies and depends on different organizations and companies. Nevertheless, every type of information needs some security to preserve data. The level of security depends on the nature of information. For example, the military databases require top and high level security so that the information is not accessed by an outsider but the concerned authority because the leakage of critical information in this case could be dangerous and even life threatening. Whereas the level of security needed for the database of a public server may not be as intensive as the military database.

Database security is essential because they suffer from security threats that may prove harmful and disastrous if disclosed or accessed publicly. Below we will present some security threats that are suffered by the databases.

- **Privilege Abuse:** When database users are provided excessive privileges than their required functionality, then these privileges can be intentionally or unintentionally exploited.
- **Legitimate Privilege Abuse:** In this attack, the attacker with the legitimate privilege access to the database may abuse the information stored in the database for the malicious purposes.
- **Privilege Promotion:** The attacker in this attack takes advantage of the software vulnerabilities and errors and then elevates his clearance level to access the critical information stored in the database.
- **Operating System Vulnerabilities:** In operating system vulnerabilities, the attacker exploits the vulnerabilities in the operating system to gain unauthorized access to the database for malicious reasons.

To alleviate these database security attacks many database security techniques have been proposed. Most of these techniques strengthen the access control mechanisms to restrict illegal access to the database.

In our paper we discuss some of the common security techniques that can be used in addition to the access control mechanisms to store the data securely and in such a way that any outsider will not be able to understand the information, even if he is able to retrieve it.

Below we list some common security schemes that can be applied to database security.

- **Cryptography:** Cryptography is the study and the practice of the techniques wherein the plaintext by encryption is converted to an obfuscated and non-readable text.
- **Hashing:** Hashing is defined as the transformation of a variable length data into a fixed length string by using hash functions without which the retrieval of the data is not possible.
- **Steganography:** Steganography is process of concealing sensitive information in any type of cover media.
- **Access Control:** Access control mechanisms restrict the access to the database and its information to outsiders except for the authorized users.

The rest of the paper is organized as follows: Section 2 illustrates some common security techniques that can be implemented in enhancing the security of the database. A Comparative Analysis is presented in Section 3 followed by the Conclusion in Section 4.

## 2. VARIOUS DATABASE SECURITY TECHNIQUES

In this section we list common security techniques that may prove useful in fortifying the database.

### 2.1. Securing Database using Cryptography

Sesay et al. proposed a database encryption scheme. In this scheme the users are divided into two levels: Level 1 (L1) and Level 2 (L2) [1], [2]. Level 1 users have access to their own private encrypted data and the unclassified public data, whereas Level 2 users have access to their own private data and also classified data which is stored in an encrypted form.

Liu et al. proposed a novel database encryption mechanism [3], [4]. The proposed mechanism performs column-wise encryption that allows the users to classify the data into sensitive data and public data. This classification helps in selecting to encrypt only that data which is critical and leaves the public data untouched thereby reducing the burden of encrypting and decrypting the whole database, as result of which the performance is not degraded.

Mixed Cryptography Database [5] scheme is presented by Kadhem et al. The technique involves designing a framework to encrypt the databases over the unsecured network in a diversified form that comprise of owning many keys by various parties. In the proposed framework, the data is grouped depending upon the ownership and on other conditions.

### 2.2. Securing Database using Steganography

Das et al. [6] explained various techniques in steganography that can be implemented to hide critical data and prevent them from unauthorized and direct access. The various techniques include still image steganography, audio steganography, video steganography, IP Datagram steganography.

Naseem et al. presented a method that uses steganography to hide data [7]. In the proposed scheme the data is embedded in the LSB's of the pixel values. The pixels values are categorized into different ranges and depending on the range certain number of bits is allocated to hide the sensitive data.

Kuo et al. presented a different approach to conceal data. In this scheme the image is divided into fixed number of blocks [8], [9], [10], [11]. Histogram of each block is calculated along with the maximum and minimum points to mask the data. This mechanism increases the hiding capacity of the data.

Dey et al. employs a diverse approach to efficiently hide the sensitive data and escalate the data hiding capacity in still images [12], [13], [14]. The technique involves using prime numbers and natural numbers to enhance the number of bit planes to cloak the data in the images.

### 2.3. Securing Database using Access Control

Bertino et al. [15] explains an authorization technique for video databases. In the proposed scheme, the access to the database and to a particular stream of the video is granted only after verifying the credentials of that user. The credentials may not just be the user-id but it may be the characteristics that define the user and only after successful verification of the credentials the user is granted the permission to access the database.

Kodali et al. presented a generalized authorization model for multimedia digital libraries [16], [17], [18]. The scheme involves integrating the three most common and widely used access control mechanisms namely: mandatory, discretionary and role-based models into a single framework to allow a unified access to the protected data. The technique also addresses the need of continuous media data while supporting the QoS constraints alongside preserving the operational semantics.

An authorization model is proposed by Rizvi et al [19], [20], [21], [22], [23]. In the explained technique is based on authorization views which enable authorization transparent querying in which the user queries are formed and represented in terms of database relations and are acceptable only when the queries can be verified using the information contained in the authorization rules. The work presents the new techniques of validity and conditional validity which is an extension of the earlier work done in the same area.

### 3. COMPARATIVE ANALYSIS OF VARIOUS DATABASE SECURITY TECHNIQUES

In [1], [3], [5] cryptography is implemented to keep the data in the database secure by encrypting the data. In [1], the authors have categorized the users in two levels: Level 1 and Level 2. Based on the Level of the user the accessibility of data provided. Level 1 users are allowed to access their own private encrypted data and the public data, whereas Level 2 user is permitted to access both, the encrypted private data and the encrypted classified data. The advantage of this scheme is that the grouping of users and grouping of data into two levels avoids the burden of unnecessary encryption of the whole data. Only the classified data and the private data are encrypted and the public data is left unchanged. A rather different approach is followed by [3]. In [3], the authors perform column-wise encryption of the data that is defined as the sensitive data by the users. The column-wise encryption approach prevents the whole database to be encrypted but only the critical information, thus averting the performance degradation problem of the database during the retrieval of the data. A varied style is undertaken by the authors in [5]. In this scheme mixed mode cryptography is employed to secure the database over the untrusted and unreliable network in a mixed form. Many keys are held by different parties who have the access to the database so that even when the database is attacked at multiple points by an insider or an outsider the database is not comprised.

The authors in [7], [8], [12] have proposed steganography as a method that can be implemented to secure the data in the database. In [7] still images are used to hide the data. In the explained scheme the pixels in the image grouped based on their intensity to hide the data. The advantage possessed by this method is that based on the intensity of the pixels a varied number of bits can be utilized to conceal the data instead of fixed number of bits. Instead of grouping the pixels of the complete image based on the intensity, the authors in [8] divide the image in equal sized blocks and the histogram of these blocks is calculated. The maximum and minimum points of the histogram are recorded and the critical data is embedded in between these points. The benefit in this approach is that the embedding capacity of the image is enhanced. To achieve the same result of increase the hiding capacity of the image and to conceal the data the authors in [12] have followed a different approach. In this technique prime numbers are used to utilize not just the lower bit planes of the image but even the higher bit planes of the image, thereby extending the hiding capacity.

Authorization techniques are presented in [15], [16], [19]. In [15], authorization techniques for video database are proposed. The scheme involves authorizing the users based on their credentials to provide access to the video database. The credentials may contain the characteristics defined for the users and not just their user identifications. A stricter approach is followed by [16]. In the method involves integrating the mandatory, discretionary and role-based models into a unified framework that grants access to the data in the database. Another technique based on access control mechanism is proposed by [19]. The authors in [19] use authorization views that enable transparent querying which are validated only when the information is present in the authorization views otherwise they are not. The benefit provided by this approach is that only the information and rules present in the authorization views are accepted and only then the access is granted otherwise the access is denied to the database.

TABLE 1 below lists a brief overview of the security techniques along with their description.

## 4. CONCLUSION

In this paper we talked about various security vulnerabilities that the database suffers from and the need for security to alleviate these vulnerabilities. We also presented some common security techniques that can be employed to augment and enhance the security of the database against some known attacks and security threats.

In Section 1 we provided an introduction about the database and the security threats and need for security in the database. In the next section we discussed various security techniques that may be implemented in the database. A Comparative Analysis of the techniques discussed in Section 2 is presented in Section 3.

## REFERENCES

[1] Samba Sesay, Zongkai Yang, Jingwen Chen, Du Xu, "A Secure Database

Table 1: A summary of the various Data Security Techniques

| Security Technique | Brief Description of the Technique | Security Techniques Proposed |
|---|---|---|
| 1. Cryptography | Cryptography is a technique to protect the data in which the plaintext or data is converted into a cipher text by applying encryption so that the resultant cipher text is in an unreadable form. | 1. In [1] the authors presented that only encrypts the classified data and the unclassified data is left unchanged. Access is granted to the data by grouping the users into two levels: Level 1 and Level 2.<br>2. The authors in this technique [3] allow the users to classify the data based on the nature of the data as critical or public. Only critical data is encrypted column-wise which helps in avoiding the degradation of the performance.<br>3. In [5] a novel approach of mixed cryptography is presented. The proposed is applied to databases that reside over the unsecured networks. |
| 2. Steganography | Steganography is technique of hiding or concealing the sensitive data in any type of cover media. | 1. The authors in [7] explained a method that focuses on utilizing the intensity of the pixels to hide data instead of the conventional approach where just 1 LSB bit is used to hide the critical data.<br>2. In the next approach a single image is divided into blocks of equal sizes [8] whose histogram is calculated and the data is hidden in the histograms of these blocks. |
| 3. Access Control/Authorization | Access Control/ authorizations mechanisms restrict the illegal access to databases and those with valid credential and the authorized users are granted access to the database. | 1. Authorization technique for video database is proposed in [15]. In the proposed scheme a user is granted access to the video stored in the database only after verifying its credentials.<br>2. In [16] a generalized |
| | | framework is presented. In this method, mandatory, discretionary, and role-based models are integrated into a unified framework to allow access to the database.<br>3. A method based on authorization views is explained in [19]. The scheme enables transparent querying in which the user's queries are represented into database relations. The users are granted access only when the information is present in authorization view. |